\documentclass[a4paper]{jpconf}
\usepackage{graphicx}
\usepackage{textgreek}
\usepackage[utf8x]{inputenc}
\begin{document}
\title{Energy dependence of hadron spectra and yields in p+p and $^7$Be+$^9$Be collisions from the NA61/SHINE experiment at the CERN SPS}

\author{D.T. Larsen for the NA61/SHINE Collaboration}

\address{Uniwersytet Jagielloński, Łojasiewicza 11, 30-348 Kraków, Poland}

\ead{dlarsen@cern.ch}

\begin{abstract}
The NA61/SHINE programme on strong interactions covers the study of the onset of deconfinement and aims to discover the critical point of strongly interacting matter by performing an energy and system-size scan over the full CERN SPS momentum range.
So far the scans of p+p, $^7$Be+$^9$Be, as well as Ar+Sc have been completed.
Results from p+p and Be+Be collisions are now emerging, in particular the energy dependence of hadron spectra and yields.
Status and preliminary results from this effort will be presented.
\end{abstract}

\section{Introduction}
NA61/SHINE~\cite{BeBe1} is a fixed target experiment at the CERN SPS.
Tracking is provided by five time projection chambers.
Additionally, particle identification is aided by time-of-flight detectors. A modular calorimeter is used to determine the collision centrality for nucleus-nucleus collisions.
Both primary and secondary beams are available to the experiment, allowing data taking with projectile sizes ranging from proton to lead, as well as with pions and kaons.
Besides studying strong interactions, the experiment also performs precise hadron production reference measurements for neutrino (Fermilab and T2K) and cosmic-ray (KASCADE and Pierre Auger Observatory) physics.

\section{Analysis of identified hadron spectra}
\begin{figure}
\begin{minipage}[b]{0.38\textwidth}
\includegraphics[trim = 0mm 0mm 0mm 0mm, clip, width=\textwidth]{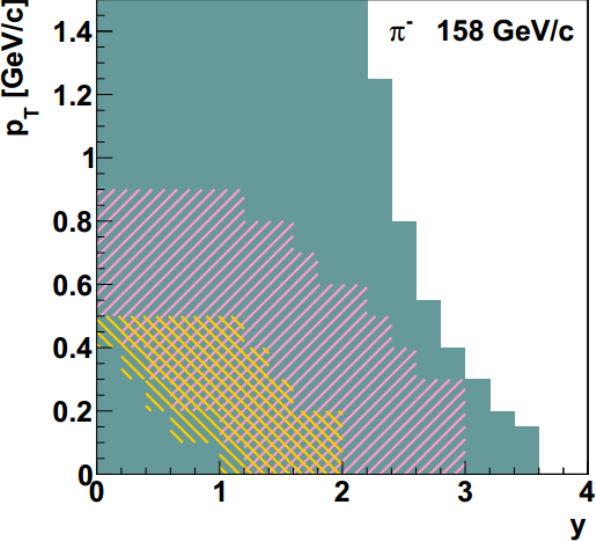}
\caption{\label{BeBe1} Acceptance for various data analysis methods. Blue, solid area: h$^-$; magenta lines: $dE/dx$; yellow lines: time-of-flight.}
\end{minipage}
\hspace{0.02\textwidth}
\begin{minipage}[b]{0.58\textwidth}
\includegraphics[trim = 0mm 2mm 0mm 5.5mm, clip, width=\textwidth]{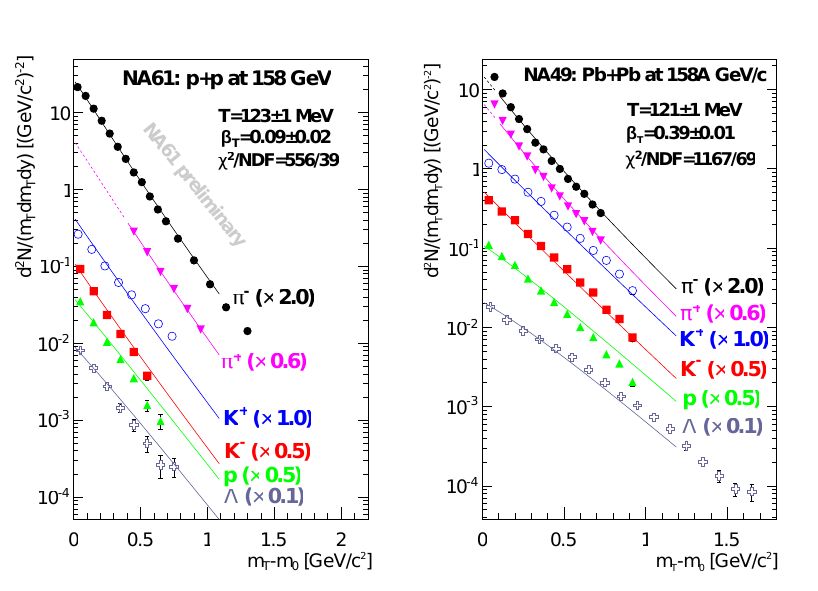}
\caption{\label{pp1} Transverse mass spectra at mid-rapidity. Left: NA61/SHINE inelastic p+p interactions at 158~GeV/c; right: NA49 central Pb+Pb collisions at 158{\it A}~GeV/c.}
\end{minipage}
\end{figure}

More than 90\% of primary negative particles produced in inelastic interactions at SPS energies are π$^-$.
Thus, the π$^-$ spectra may be obtained by subtracting the small non-pion contribution from the overall spectra for negatively charged hadrons.
This non-pion contribution is taken from the EPOS model~\cite{pp2}.
Since particle identification is not required, a very large phase-space acceptance is obtained.
In addition to the h$^-$ method~\cite{pp1}, $dE/dx$ and time-of-flight may be used for proper identification.
The $dE/dx$ method is based on the particle energy loss in the TPC gas, while the particle mass 
can be identified using the time-of-flight information.
The acceptances for these methods are shown in \Fref{BeBe1} for p+p interactions at 158~GeV/c.
The results are corrected for detector inefficiencies, feed-down from weak decays and secondary interactions, contribution from non-target interactions, as well as trigger and event selection biases.

\section{p+p results}
The results presented below were obtained for identified hadrons produced in inelastic p+p interactions at 20, 31, 40, 80 and 158~GeV/c~\cite{szymon}.

\Fref{pp1} shows spectra of transverse mass of negatively and positively charged π, K, p and Λ produced in inelastic p+p interactions at mid-rapidity.
Corresponding NA49 measurements~\cite{pp3,pp4,pp6} for central Pb+Pb collisions are also shown.
The data was fitted using a blast wave model parametrisation~\cite{pp7} $\frac{dN_i}{m_Tdm_Tdy}=A_im_TK_1\left(\frac{m_T\cosh\rho}{T}\right)I_0\left(\frac{p_T\sinh\rho}{T}\right)$.
The parameter $\rho$ is related to the transverse flow velocity $\beta_T$ by $\rho=\tanh^{-1}\beta_T$.
One finds that $\beta_T$ is significantly smaller in p+p than Pb+Pb collisions.
While the spectra are approximately exponential in p+p reactions, this exponential dependence is modified in Pb+Pb interactions by the transverse flow.

\begin{figure}
\includegraphics[trim = 0mm 1mm 0mm 1mm, clip, width=0.3333\textwidth]{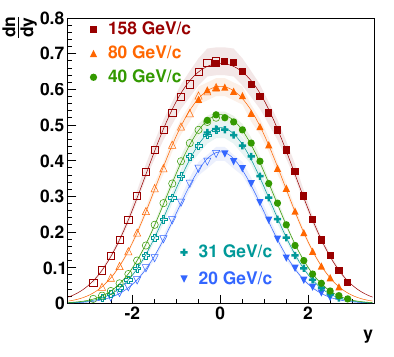}
\includegraphics[trim = 0mm 1mm 0mm 1mm, clip, width=0.3333\textwidth]{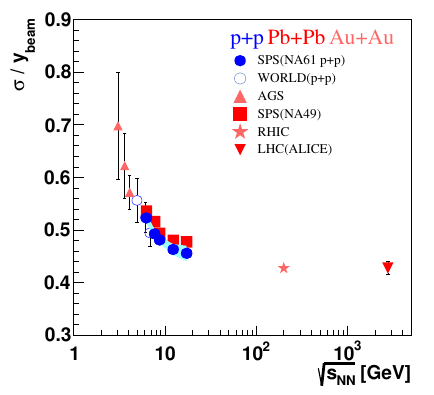}
\includegraphics[trim = 0mm 1mm 0mm 1mm, clip, width=0.3333\textwidth]{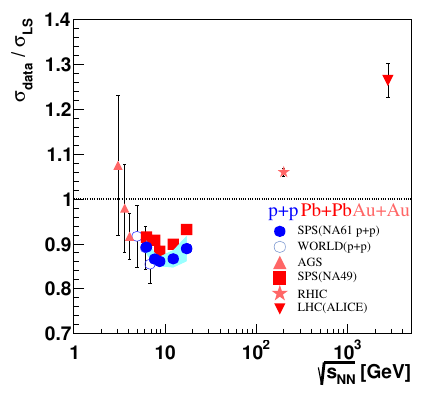}
\caption{\label{pp2} Left: π$^-$ rapidity spectra in inelastic p+p interactions. Centre: energy dependence of the scaled width of rapidity spectra. Right: energy dependence of the ratio of the measured width to that predicted by the hydro-dynamical model~\cite{pp8,pp9}. World data from Refs.~\cite{pp10,pp11,pp12}. Not corrected for isospin effects.}
\end{figure}

\Fref{pp2} shows rapidity spectra of π$^-$ produced in inelastic p+p interactions (left), the energy dependence of the rms width $\sigma$ of those spectra divided by beam rapidity (centre) and divided by predictions ($\sigma_{LS}$) from the hydro-dynamical model~\cite{pp8,pp9} (right).
Although the shapes of the rapidity spectra are approximately Gaussian, the best fit was obtained using a sum of two Gaussians.
It can be seen in \Fref{pp2}~(centre) that $\sigma/y_{beam}$ for π$^-$ decreases with increasing collision energy.
While $\sigma/\sigma_{LS}$ and $\sigma/y_{beam}$ are smaller in p+p than Pb+Pb collisions, no qualitative difference is observed for their energy dependence.

\begin{figure}
\includegraphics[trim = 0mm 0.5mm 0mm 2mm, clip, width=0.6666\textwidth]{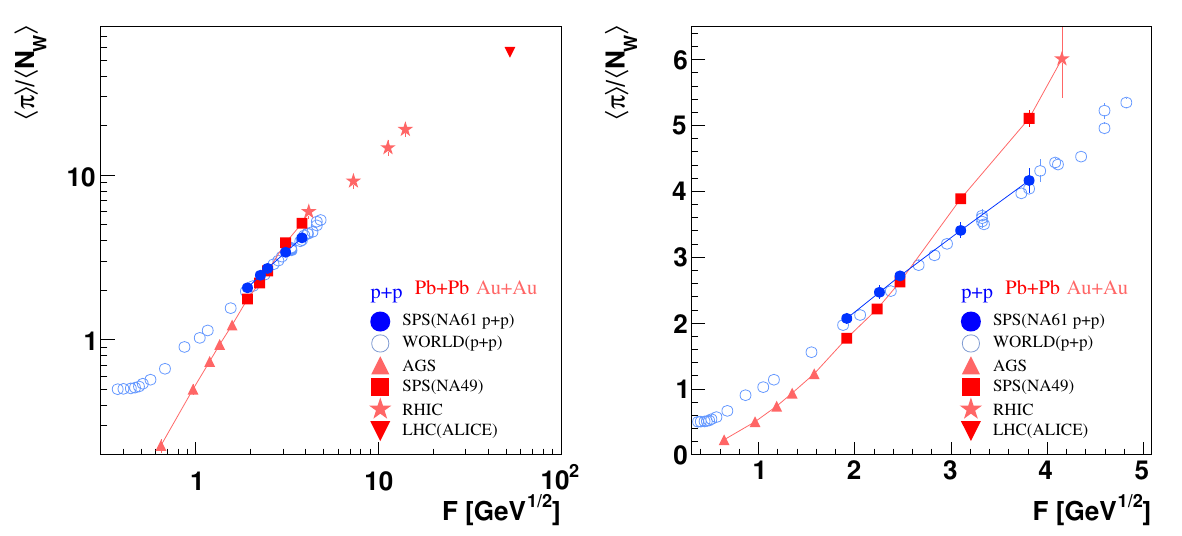}
\begin{minipage}[b]{0.3333\textwidth}
\caption{\label{pp3} Energy dependence of mean π multiplicity per wounded nucleon in inelastic p+p interactions and central Pb+Pb (Au+Au) collisions. World data from Refs.~\cite{pp11,pp13}.}
\end{minipage}
\end{figure}

The ``kink'', ``horn'' and ``step''~\cite{pp4,pp5} are important signals for the onset of deconfinement observed in central Pb+Pb interactions.
\Fref{pp3} shows that the energy dependence of mean π multiplicity increases slower in p+p than in Pb+Pb collisions (``kink'').
Hence, the two dependences cross each other at around 40{\it A}~GeV/c.

\begin{figure}
\includegraphics[trim = 6mm 3mm 5mm 5.5mm, clip, width=0.6666\textwidth]{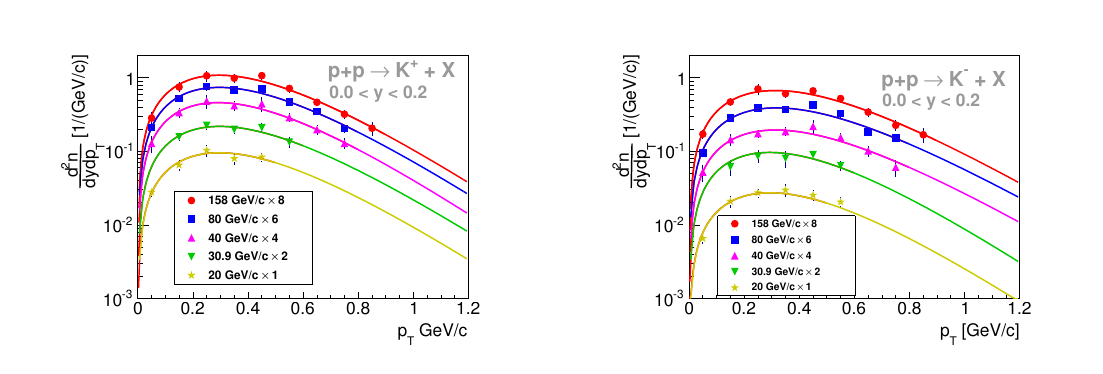}
\begin{minipage}[b]{0.3333\textwidth}
\caption{\label{pp4} Transverse momentum spectra of K mesons at mid-rapidity in inelastic p+p interactions at CERN SPS energies.}
\end{minipage}
\end{figure}

The new measurements of NA61/SHINE significantly improve the world data for the inverse-slope parameter $T$ of $m_T$ spectra of Kaons~\cite{pp14} and for yields of π$^+$ mesons at mid-rapidity~\cite{pp15,pp16}.
The $m_T$ spectra were fitted to the exponential function $\frac{d^2n}{dp_Tdy}=\frac{Sp_T}{T^2+m_KT}\exp\left(-\frac{\sqrt{p^2_T+m^2_K}-m_K}{T}\right)$.
$S$ is the yield integral and $m_K$ is the K mass.
\Fref{pp4} shows the result of the fit to the spectra of K$^+$ and K$^-$ at mid-rapidity.

\begin{figure}
\includegraphics[trim = 0mm 3mm 0mm 1.5mm, clip, width=0.6666\textwidth]{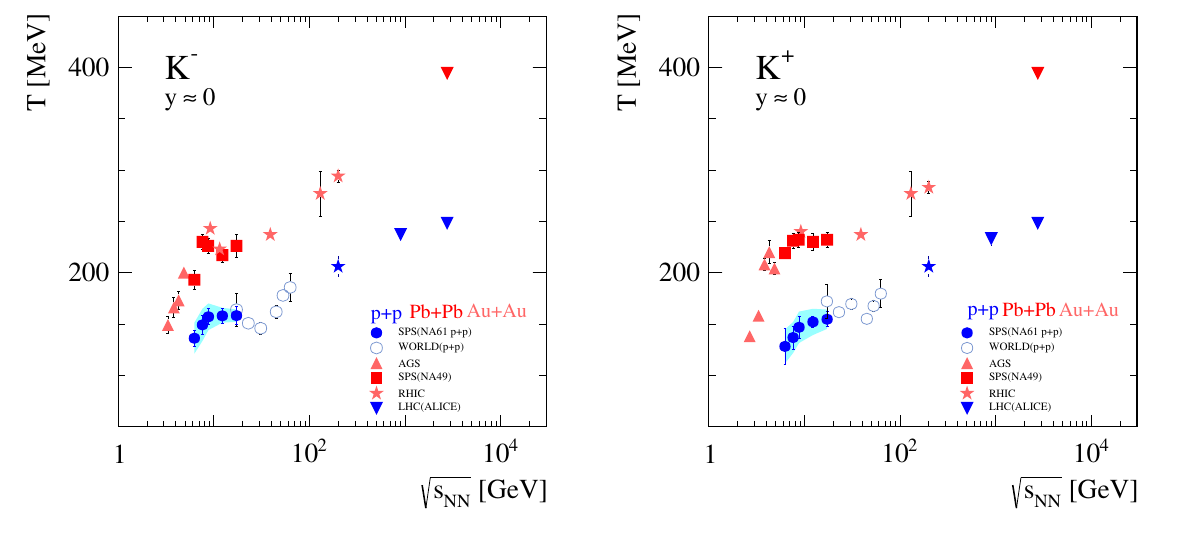}
\begin{minipage}[b]{0.3333\textwidth}
\caption{\label{pp5} Energy dependence of the inverse slope parameter T of transverse mass spectra of K$^+$ and K$^-$ in inelastic p+p and central Pb+Pb/Au+Au interactions. World data from Refs.~\cite{pp14, pp17, pp18, pp19}.}
\end{minipage}
\end{figure}

\Fref{pp5} shows the energy dependence of $T$ for K$^+$ and K$^-$.
Surprisingly, the results from inelastic p+p interactions exhibit rapid changes similar to those observed in central Pb+Pb collisions (``step'').

\begin{figure}
\includegraphics[trim = 0mm 3.5mm 0mm 1.5mm, clip, width=0.3333\textwidth]{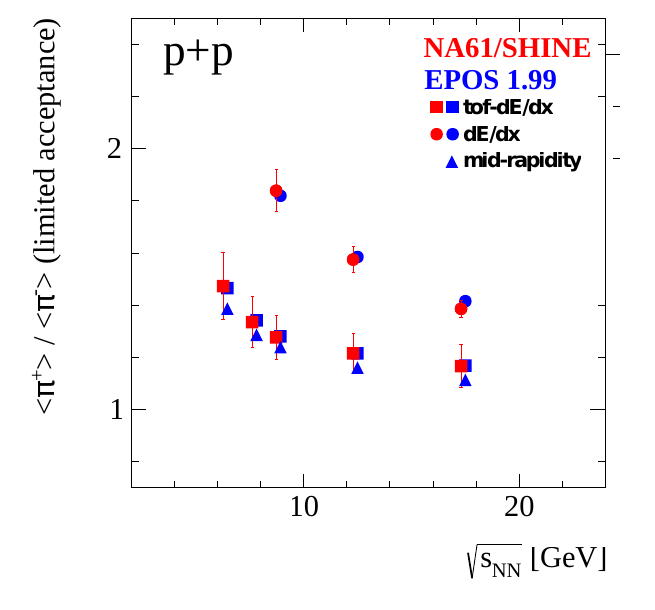}
\includegraphics[trim = 0mm 2.5mm 0mm 1.5mm, clip, width=0.3333\textwidth]{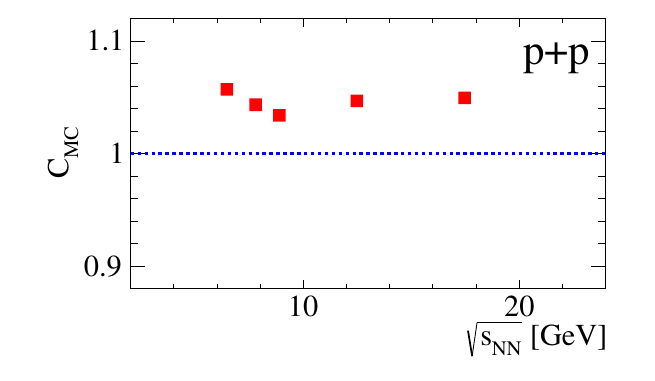}
\begin{minipage}[b]{0.3333\textwidth}
\caption{\label{pp6} Left: energy dependency of π$^+$/π$^-$ ratio in inelastic p+p interactions compared with EPOS predictions. Right: correction factors used to obtain π$^+$ yields from π$^-$ yields.}
\end{minipage}
\end{figure}

While the h$^-$ method~\cite{pp1} can be used to measure spectra for π$^-$, a similar approach cannot be used for π$^+$ because of the large K and p contamination.
Instead, the ratio of the measured π$^+$ and π$^-$ yields was calculated within the acceptance of the time-of-flight and/or $dE/dx$ methods, and compared with model predictions.
As can be seen in \Fref{pp6}~(left), the agreement between the measurements and the EPOS model~\cite{pp2} is better than 0.1\%.
Finally, the π$^+$ mid-rapidity yield was calculated as the product of the measured π$^-$ yield at $y=0$, the measured π$^+$/π$^-$ ratio within the time-of-flight and $dE/dx$ acceptance, and the EPOS correction factor $C_{MC}$ (see \Fref{pp6}~(right): $\pi^+(y=0)=\pi^-(y=0)\pi^+/\pi^-(tof-dE/dx)C_{MC}$, where $C_{MC}=\left[\frac{\pi^+/\pi^-(y=0)}{\pi^+/\pi^-(tof)}\right]_{MC}\approx 5\%$.

\begin{figure}
\includegraphics[trim = 0mm 3mm 0mm 1.5mm, clip, width=0.6666\textwidth]{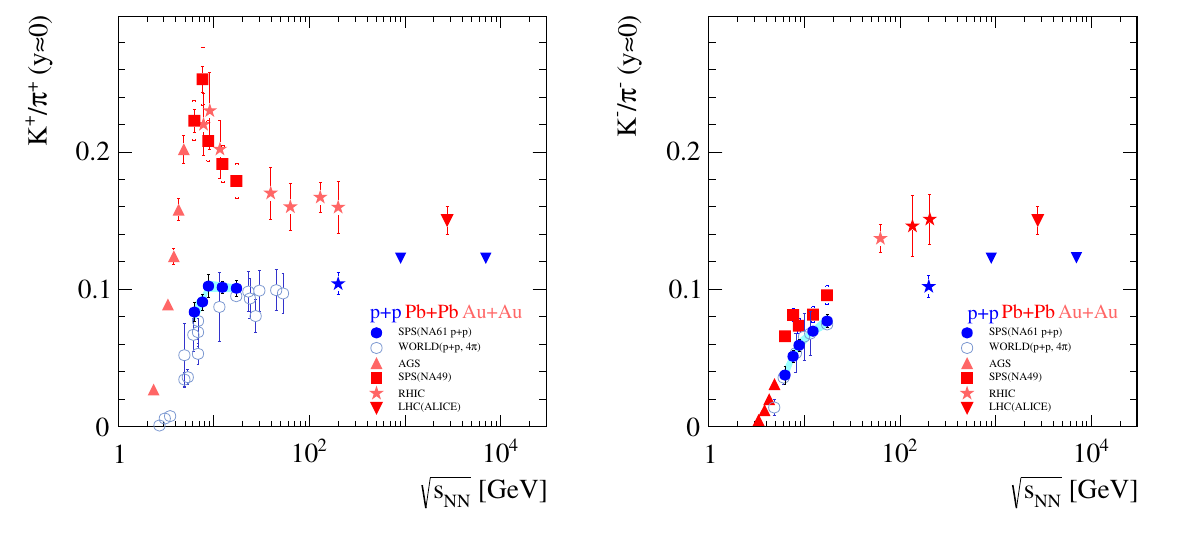}
\begin{minipage}[b]{0.3333\textwidth}
\caption{\label{pp7} Energy dependence of the K$^+$/π$^+$ and K$^-$/π$^-$ ratios in inelastic p+p and central Pb+Pb and Au+Au interactions. World data from Refs.~\cite{pp13, pp15, pp16, pp19, pp20}.}
\end{minipage}
\end{figure}

\Fref{pp7} shows the energy dependence of the K$^+$/π$^+$ and K$^-$/π$^-$ ratios observed in inelastic p+p interactions, as well as world data for central Pb+Pb and Au+Au interactions as reference.
Again surprisingly, the data suggests that even inelastic p+p interactions exhibit a step structure in the energy dependence of the K$^+$/π$^+$ ratio, which may be considered a precursor of the ``horn'' found in central  Pb+Pb collisions.

\begin{figure}
\includegraphics[trim = 0mm 3mm 0mm 1.5mm, clip, width=0.6666\textwidth]{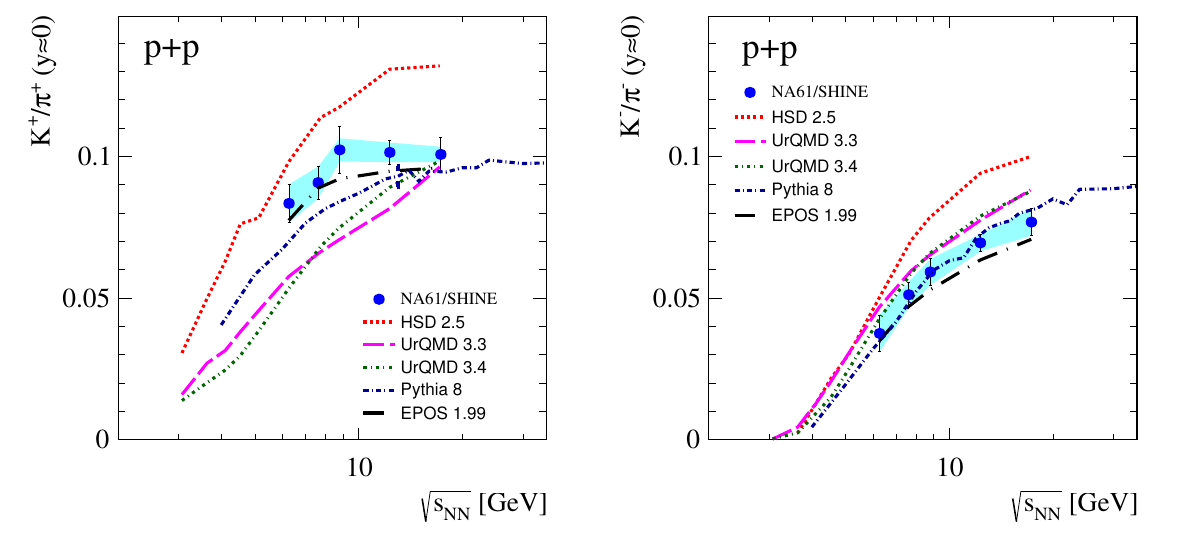}
\begin{minipage}[b]{0.3333\textwidth}
\caption{\label{pp8} Energy dependence of the K$^+$/π$^+$ and K$^-$/π$^-$ ratios in inelastic p+p interactions compared~\cite{pp25} with theoretical models EPOS~\cite{pp2}, UrQMD~\cite{pp22,pp21}, Pythia 8~\cite{pp24} and HSD~\cite{pp23}.}
\end{minipage}
\end{figure}

Furthermore, the measured ratios were compared to theoretical models, as shown in \Fref{pp8}.
It is evident that none of the models adequately describes the data.

\begin{figure}
\includegraphics[trim = 0mm 3mm 5mm 5mm, clip, width=0.4000\textwidth]{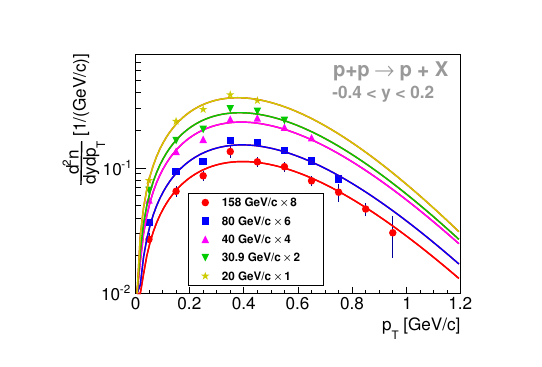}
\includegraphics[trim = 0mm 2mm 3mm 1mm, clip, width=0.2666\textwidth]{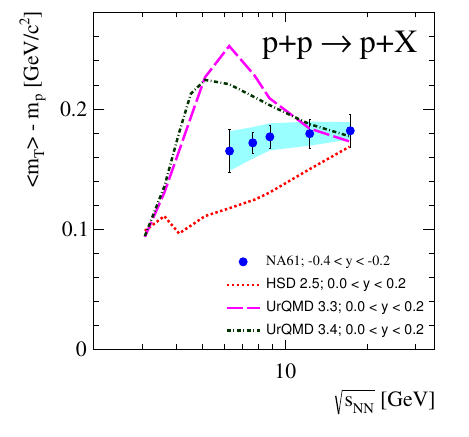}
\begin{minipage}[b]{0.3333\textwidth}
\caption{\label{pp9} Left: transverse momentum spectra of protons at mid-rapidity in inelastic p+p interactions. Right: comparison of $\langle m_T\rangle-m_p$ calculated using the spectra from the left plot with predictions of models~\cite{pp25}.}
\end{minipage}
\end{figure}

\Fref{pp9} shows for protons from inelastic p+p interactions the spectra of $p_T$ (left) and $\langle m_T \rangle$ (right) as a function of collision energy.
Neither the magnitude nor the energy dependence of $\langle m_T \rangle$ is reproduced by the UrQMD or the HSD model.

\section{Be+Be results}
\begin{figure}
\includegraphics[trim = 0mm 1mm 0mm 3.5mm, clip, width=0.5\textwidth]{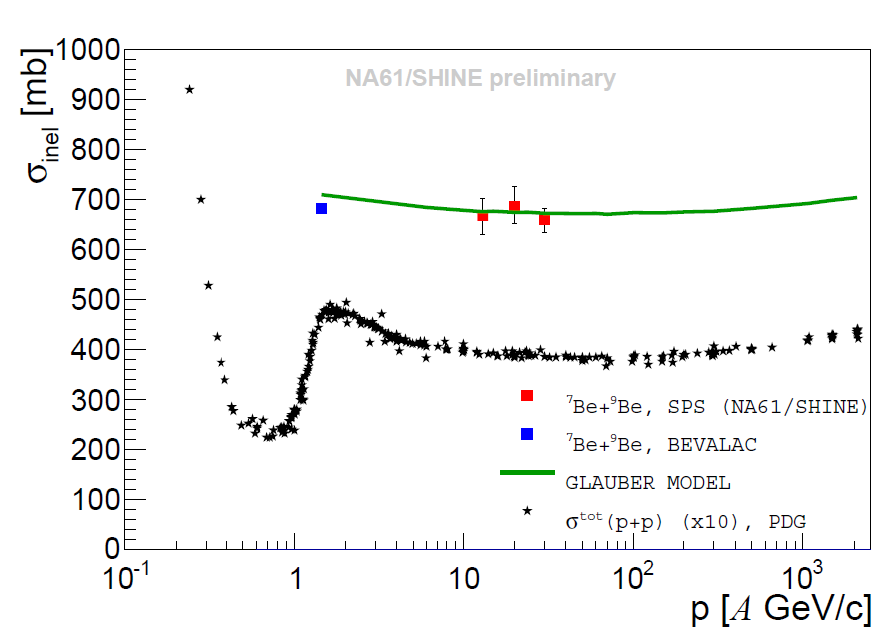}
\begin{minipage}[b]{0.5\textwidth}
\caption{\label{BeBe5} Inelastic cross section of $^7$Be+$^9$Be collisions as well as total cross section for p+p interactions as a function of beam momentum. The curve shows the prediction of the Glauber model.}
\end{minipage}
\end{figure}

Previously, the inelastic cross section for $^7$Be+$^9$Be was only measured at 1.45{\it A}~GeV/c~\cite{BeBe5}.
The new measurements of NA61/SHINE now extend this to 13{\it A}, 20{\it A} and 30{\it A}~GeV/c~\cite{BeBe6}.
A scintillator counter placed behind the target measures the square of the charge of a particle passing through it.
Inelastic interactions were selected by requiring a signal below that expected for an intact beam nucleus.
Data was also taken with the target removed to be able to subtract the background caused by beam interactions with detector material.
The interaction probability is given by $P_{int}=\frac{P_I-P_R}{1-P_R}$, where $P_I$ and $P_R$ are the interaction probabilities when the target is inserted and removed, respectively.
Using $P_{int}$, the cross section can be calculated from $\sigma_{inel}=\frac{1}{\rho L_{eff} N_A/A}P_{int}$, $L_{eff}=\lambda_{abs}(1-e^{-L/\lambda_{abs}})$ and $\lambda_{abs}=\frac{A}{\rho N_A\sigma_{inel}}$.
$N_A$, $\rho$, $L$ and $A$ are the Avogadro constant, the target density, length and atomic number, respectively.
\Fref{BeBe5} shows that the new measurements are in agreement with the previous measurement as well as the Glauber model~\cite{BeBe7} prediction.

\begin{figure}
\includegraphics[trim = 0mm 3mm 0mm 5mm, clip, width=0.6666\textwidth]{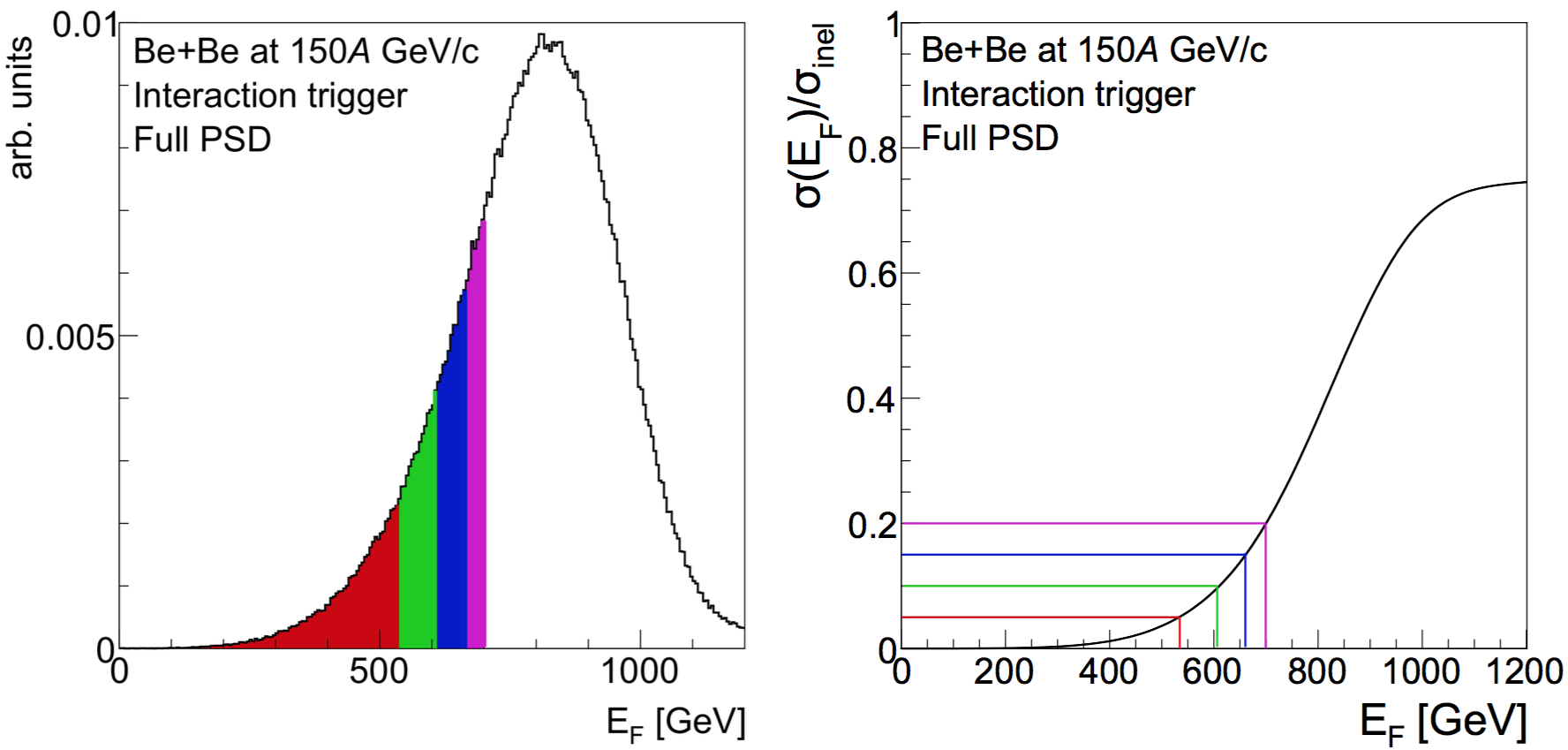}
\begin{minipage}[b]{0.3333\textwidth}
\caption{\label{BeBe6} Left: distribution of forward energy $E_F$ measured in the PSD calorimeter in $^7$Be+$^9$Be collisions at 150{\it A}~GeV/c. Right: definition of forward energy event classes.}
\end{minipage}
\end{figure}

Preliminary results on spectra were obtained for $^7$Be+$^9$Be interactions at beam momenta of 20{\it A}, 30{\it A}, 40{\it A}, 75{\it A} and 150{\it A}~GeV/c in the centrality classes 0--5\%, 5--10\%, 10--15\% and 15--20\%~\cite{emil}.
Unless otherwise stated, only statistical errors are shown.
The centrality was derived from the energy deposited in the forward calorimeter PSD.
\Fref{BeBe6} shows the relationship between deposited energy and the centrality classes.
Due to the modularity of this detector, it is possible to change the acceptance during off-line analysis.
A smaller acceptance will cause some of the projectile spectators to be lost, but at the same time reduces the contribution from particles produced in the collision.
Thus, the result will depend on the selected acceptance.
This effect is largest for low beam momenta.
For example for the total π$^-$ multiplicity at 20{\it A}~GeV/c, the spread for different acceptances is up to 5\%.

\begin{figure}
\includegraphics[trim = 0mm 0.5mm 0mm 1mm, clip, width=0.3333\textwidth]{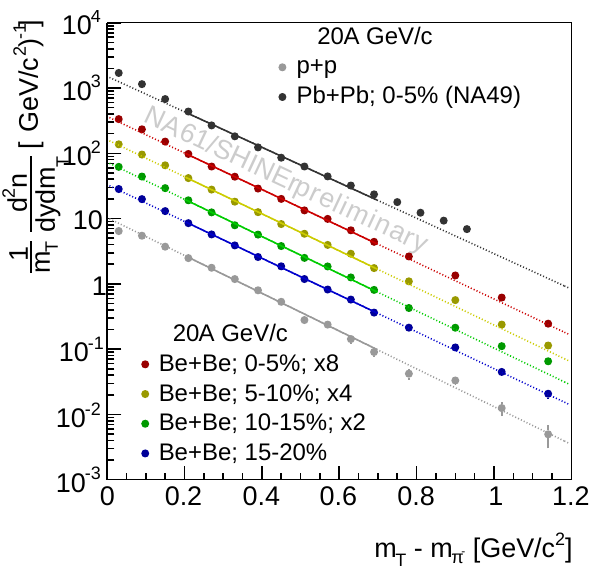}
\includegraphics[trim = 0mm 0.5mm 0mm 1mm, clip, width=0.3333\textwidth]{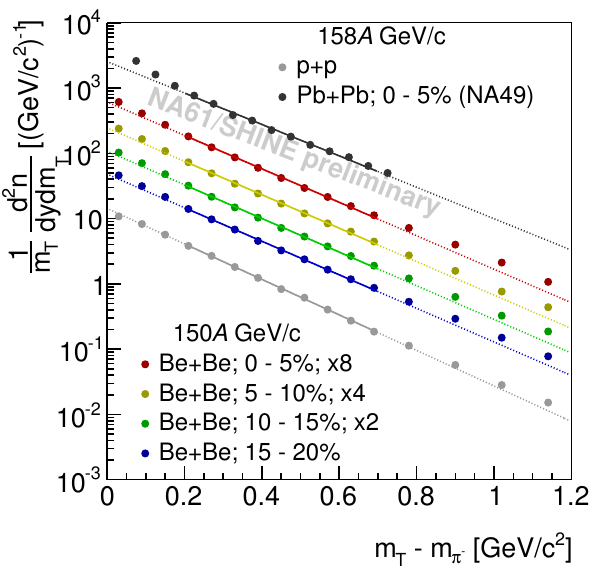}
\begin{minipage}[b]{0.3333\textwidth}
\caption{\label{BeBe10} Transverse mass spectra of π$^-$ mesons for $^7$Be+$^9$Be, p+p and Pb+Pb interactions. Left: 20{\it A}~GeV/c; right: 150{\it A}~GeV/c beam momentum.}
\end{minipage}
\end{figure}

\Fref{BeBe10} shows for $^7$Be+$^9$Be π$^-$ collisions the $m_T$ spectra for the different centrality classes for the beam momenta 20{\it A} and 150{\it A}~GeV/c, as well as for inelastic p+p and central Pb+Pb interactions.
The p+p data is described very well by an exponential function, while both $^7$Be+$^9$Be and Pb+Pb spectra deviate from this function at low and high values of $m_T$.

\begin{figure}
\includegraphics[trim = 0mm 1.5mm 0mm 1.5mm, clip, width=0.3333\textwidth]{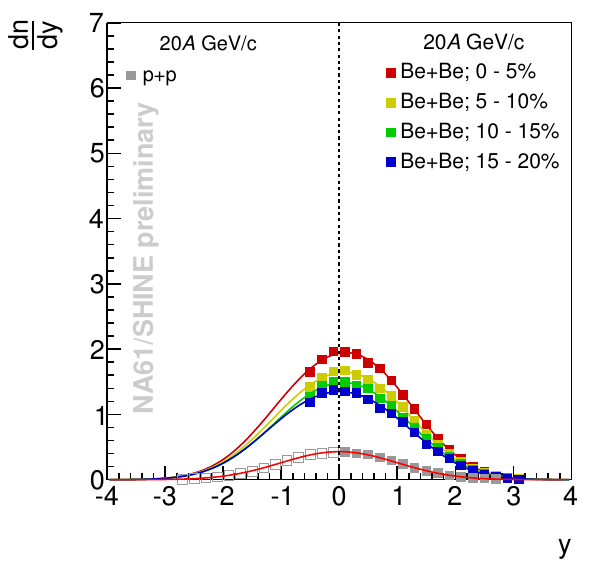}
\includegraphics[trim = 0mm 1.5mm 0mm 1.5mm, clip, width=0.3333\textwidth]{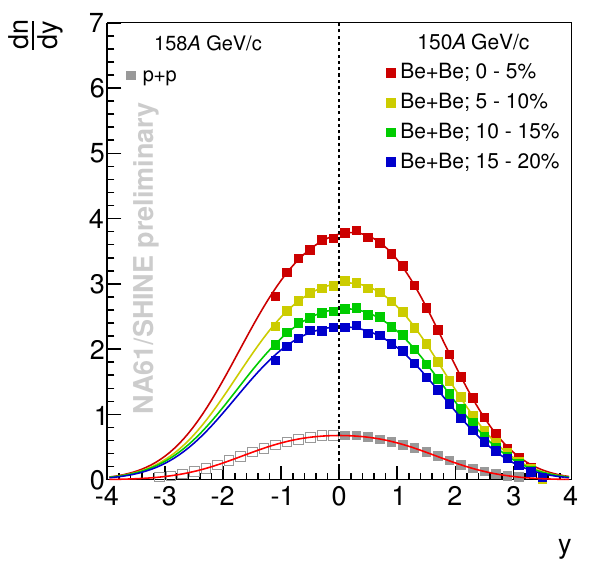}
\begin{minipage}[b]{0.3333\textwidth}
\caption{\label{BeBe7} Rapidity spectra of π$^-$ mesons for $^7$Be+$^9$Be and inelastic p+p interactions. Left: 20{\it A}~GeV/c; right: 150{\it A}~GeV/c.}
\end{minipage}
\end{figure}

Calculating normalised π$^-$ $m_T$ spectra as the ratios ($^7$Be+$^9$Be)/(p+p) and (Pb+Pb)/(p+p) facilitates the comparison of the shapes of the spectra produced in these different size systems.
\Fref{BeBe7} shows qualitatively similar behaviour for both $^7$Be+$^9$Be and Pb+Pb reactions.
The high-$m_T$ regions of both $^7$Be+$^9$Be and Pb+Pb exhibit an increase of the ratio, while for the intermediate regions a decrease is seen.
This effect is stronger for central Pb+Pb collisions.
Usually, this is attributed to collective flow.
Also, for $^7$Be+$^9$Be reactions the increase of the ratio at high values of $m_T$ appears to be larger at higher beam momenta.
This may suggest an increase of the magnitude of collective effects in $^7$Be+$^9$Be collisions with increasing beam momentum.
The low-$m_T$ regions of both $^7$Be+$^9$Be and Pb+Pb reactions exhibit an increase of the ratio.
Possible explanations include isospin asymmetry of p+p data or electromagnetic effects.
\begin{figure}
\includegraphics[trim = 0mm 1mm 0mm 1.5mm, clip, width=0.6666\textwidth]{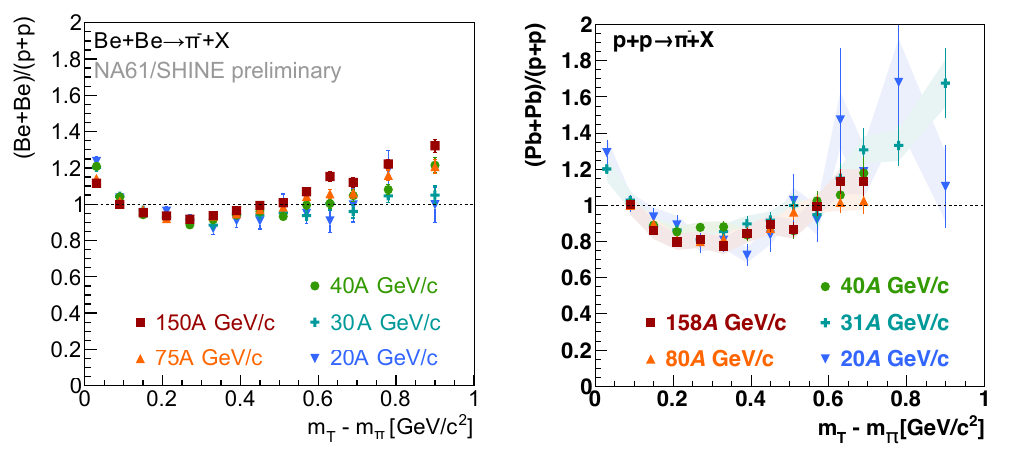}
\begin{minipage}[b]{0.3333\textwidth}
\caption{\label{BeBe11} Ratio of normalised transverse mass spectra of π$^-$ mesons for different system sizes. Left: ($^7$Be+$^9$Be)/(p+p); right: (Pb+Pb)/(p+p).}
\end{minipage}
\end{figure}

\Fref{BeBe7} shows the π$^-$ rapidity spectra from $^7$Be+$^9$Be collisions in the different centrality classes for beam momenta 20{\it A} and 150{\it A}~GeV/c, as well as for inelastic p+p reactions.
The data were fitted to two Gaussians symmetrically displaced with respect to mid-rapidity.
Although their widths are the same, the amplitudes are different due to the asymmetry of $^7$Be+$^9$Be collisions.
By extending the fit range into the backward rapidity region, it was possible to obtain a stable fit.

\begin{figure}
\includegraphics[trim = 0mm 1mm 0mm 1.5mm, clip, width=0.3333\textwidth]{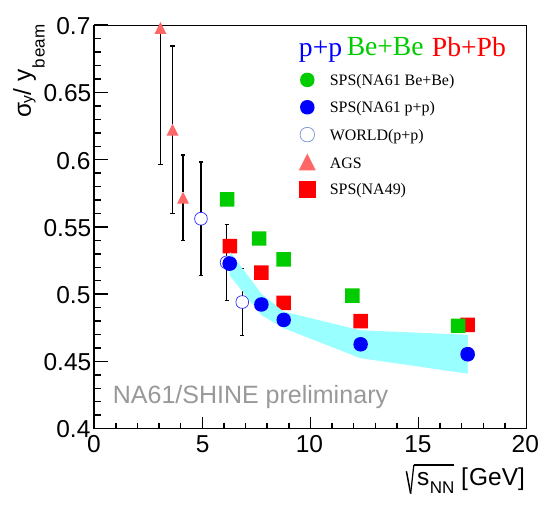}
\includegraphics[trim = 0mm 1mm 0mm 1.5mm, clip, width=0.3333\textwidth]{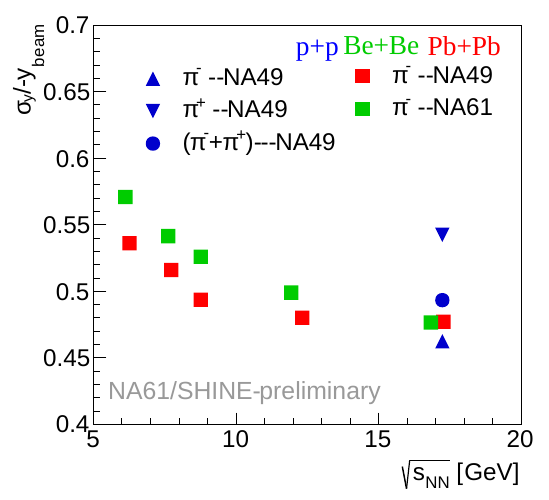}
\begin{minipage}[b]{0.3333\textwidth}
\caption{\label{BeBe89} Left: energy dependence of scaled width of $^7$Be+$^9$Be π$^-$ rapidity spectra; right: effect of isospin asymmetry in p+p interactions. World data from Refs.~\cite{pp6}.} 
\end{minipage}
\end{figure}

The rms width $\sigma_y$ of the spectra can be obtained from the fitted function.
\Fref{BeBe89} (left) shows the $\sigma_y/y_{beam}$ for $^7$Be+$^9$Be, as well as inelastic p+p and central Pb+Pb interactions.
The widths of the spectra for all systems decrease monotonically with respect to collision energy.
However, the widths of the spectra do not behave monotonically with respect to the system size for a given collision energy.
The width of the spectra from Pb+Pb lies between those of p+p and $^7$Be+$^9$Be collisions.
When comparing the rapidity spectra from different systems, one must consider that p+p has larger isospin asymmetry than $^7$Be+$^9$Be and Pb+Pb.
This isospin asymmetry can be accounted for by calculating the average spectra for π$^-$ and π$^+$.
The only large-acceptance data available for π$^+$ production in p+p in the relevant energy domain is from NA49 at 158~GeV/c~\cite{BeBe9}.
Unfortunately, this is also the energy with the smallest differences for the rapidity widths of the different system sizes.
Still, \Fref{BeBe89} (right) shows for 158~GeV/c beam momentum that by taking into account the different isospin asymmetries, the dependence of the rapidity width with respect to system size becomes monotonic.

\section{Conclusion}
The analysis of inelastic p+p and $^7$Be+$^9$Be interactions at CERN SPS energies showed many interesting effects.
In particular, the p+p data exhibited step-like structures in the energy region where the onset of deconfinement was found in central Pb+Pb collisions.
Also, many of the measurements could not be explained well by theoretical models.
For $^7$Be+$^9$Be reactions new results on the inelastic cross section were obtained at several energies. Transverse mass and rapidity spectra were measureed in the SPS energy range for three centrality classes. An indication of a transverse flow effect was found at the highest beam momenta.

\section*{Acknowledgements}
Supported by the National Science Centre of Poland (grants UMO-2014/13/N/ST2/02565, UMO-2014/12/T/ST2/00692, UMO-2013/11/N/ST2/03879 and UMO-2012/04/M/ST2/00816).

\section*{References}
\bibliography{na61References}

\providecommand{\href}[2]{#2}\begingroup\raggedright\begin{thebibliography}{10}

\bibitem{BeBe1}
N.~Abgrall {\em et~al.}, {[NA61/SHINE} Collab.]
  \href{http://dx.doi.org/10.1088/1748-0221/9/06/P06005}{{\em JINST} {\bfseries
  9} (2014) P06005},
\href{http://arxiv.org/abs/1401.4699}{{\ttfamily arXiv:1401.4699
  [physics.ins-det]}}.

\bibitem{pp2}
T.~Pierog and K.~Werner
  \href{http://dx.doi.org/10.1016/j.nuclphysbps.2009.09.017}{{\em
  Nucl.Phys.Proc.Suppl.} {\bfseries 196} (2009) 102--105},
\href{http://arxiv.org/abs/0905.1198}{{\ttfamily arXiv:0905.1198 [hep-ph]}}.

\bibitem{pp1}
N.~Abgrall {\em et~al.}, {[NA61/SHINE} Collab.]
  \href{http://dx.doi.org/10.1140/epjc/s10052-014-2794-6}{{\em Eur.Phys.J.}
  {\bfseries C74} (2014) 2794},
\href{http://arxiv.org/abs/1310.2417}{{\ttfamily arXiv:1310.2417 [hep-ex]}}.

\bibitem{szymon}
S.~Puławski, {[NA61} Collab.]
\href{http://arxiv.org/abs/1502.07916}{{\ttfamily arXiv:1502.07916 [nucl-ex]}}.

\bibitem{pp3}
C.~Alt {\em et~al.}, {[NA49} Collab.]
\href{http://dx.doi.org/10.1103/PhysRevC.73.044910}{{\em Phys.Rev.} {\bfseries
  C73} (2006) 044910}.

\bibitem{pp4}
C.~Alt {\em et~al.}, {[NA49} Collab.]
\href{http://dx.doi.org/10.1103/PhysRevC.77.024903}{{\em Phys.\ Rev.}
  {\bfseries C77} (2008) 024903}.

\bibitem{pp6}
S.~Afanasiev {\em et~al.}, {[NA49} Collab.]
\href{http://dx.doi.org/10.1103/PhysRevC.66.054902}{{\em Phys.\ Rev.}
  {\bfseries C66} (2002) 054902}.

\bibitem{pp7}
E.~Schnedermann, J.~Sollfrank, and U.~W. Heinz
  \href{http://dx.doi.org/10.1103/PhysRevC.48.2462}{{\em Phys.Rev.} {\bfseries
  C48} (1993) 2462--2475},
\href{http://arxiv.org/abs/nucl-th/9307020}{{\ttfamily arXiv:nucl-th/9307020
  [nucl-th]}}.

\bibitem{pp8}
L.~Landau
{\em Izv.Akad.Nauk Ser.Fiz.} {\bfseries 17} (1953) 51--64.

\bibitem{pp9}
E.~V. Shuryak
{\em IYF-75-4} (1974) .

\bibitem{pp10}
J.~Klay {\em et~al.}, {[E895} Collab.]
\href{http://dx.doi.org/10.1103/PhysRevC.68.054905}{{\em Phys.\ Rev.}
  {\bfseries C68} (2003) 054905}.

\bibitem{pp11}
E.~Abbas {\em et~al.}, {[ALICE} Collab.]
  \href{http://dx.doi.org/10.1016/j.physletb.2013.09.022}{{\em Phys.Lett.}
  {\bfseries B726} (2013) 610--622},
\href{http://arxiv.org/abs/1304.0347}{{\ttfamily arXiv:1304.0347 [nucl-ex]}}.

\bibitem{pp12}
S.~Adler {\em et~al.}, {[PHENIX} Collab.]
\href{http://dx.doi.org/10.1103/PhysRevC.69.034909}{{\em Phys.Rev.} {\bfseries
  C69} (2004) 034909}.

\bibitem{pp5}
M.~Gazdzicki, M.~Gorenstein, and P.~Seyboth
  \href{http://dx.doi.org/10.5506/APhysPolB.42.307}{{\em Acta Phys.Polon.}
  {\bfseries B42} (2011) 307--351},
\href{http://arxiv.org/abs/1006.1765}{{\ttfamily arXiv:1006.1765 [hep-ph]}}.

\bibitem{pp13}
B.~Abelev {\em et~al.}, {[ALICE} Collab.]
  \href{http://dx.doi.org/10.1103/PhysRevLett.109.252301}{{\em Phys.Rev.Lett.}
  {\bfseries 109} (2012) 252301},
\href{http://arxiv.org/abs/1208.1974}{{\ttfamily arXiv:1208.1974 [hep-ex]}}.

\bibitem{pp14}
M.~Kliemant, B.~Lungwitz, and M.~Gazdzicki
  \href{http://dx.doi.org/10.1103/PhysRevC.69.044903}{{\em Phys.Rev.}
  {\bfseries C69} (2004) 044903},
\href{http://arxiv.org/abs/hep-ex/0308002}{{\ttfamily arXiv:hep-ex/0308002
  [hep-ex]}}.

\bibitem{pp15}
M.~Gazdzicki and D.~Roehrich
\href{http://dx.doi.org/10.1007/BF01571878}{{\em Z.Phys.} {\bfseries C65}
  (1995) 215}.

\bibitem{pp16}
M.~Gazdzicki and D.~Rohrich \href{http://dx.doi.org/10.1007/s002880050147}{{\em
  Z.Phys.} {\bfseries C71} (1996) 55--64},
\href{http://arxiv.org/abs/hep-ex/9607004}{{\ttfamily arXiv:hep-ex/9607004
  [hep-ex]}}.

\bibitem{pp17}
B.~Abelev {\em et~al.}, {[STAR} Collab.]
  \href{http://dx.doi.org/10.1103/PhysRevC.79.034909}{{\em Phys.Rev.}
  {\bfseries C79} (2009) 034909},
\href{http://arxiv.org/abs/0808.2041}{{\ttfamily arXiv:0808.2041 [nucl-ex]}}.

\bibitem{pp18}
B.~B. Abelev {\em et~al.}, {[ALICE} Collab.]
  \href{http://dx.doi.org/10.1016/j.physletb.2014.07.011}{{\em Phys.Lett.}
  {\bfseries B736} (2014) 196--207},
\href{http://arxiv.org/abs/1401.1250}{{\ttfamily arXiv:1401.1250 [nucl-ex]}}.

\bibitem{pp19}
K.~Aamodt {\em et~al.}, {[ALICE} Collab.]
  \href{http://dx.doi.org/10.1140/epjc/s10052-011-1655-9}{{\em Eur.Phys.J.}
  {\bfseries C71} (2011) 1655},
\href{http://arxiv.org/abs/1101.4110}{{\ttfamily arXiv:1101.4110 [hep-ex]}}.

\bibitem{pp20}
I.~Arsene {\em et~al.}, {[BRAHMS} Collab.]
  \href{http://dx.doi.org/10.1103/PhysRevC.72.014908}{{\em Phys.Rev.}
  {\bfseries C72} (2005) 014908},
\href{http://arxiv.org/abs/nucl-ex/0503010}{{\ttfamily arXiv:nucl-ex/0503010
  [nucl-ex]}}.

\bibitem{pp25}
V.~Y. Vovchenko, D.~Anchishkin, and M.~Gorenstein
  \href{http://dx.doi.org/10.1103/PhysRevC.90.024916}{{\em Phys.Rev.}
  {\bfseries C90} no.~2, (2014) 024916},
\href{http://arxiv.org/abs/1407.0629}{{\ttfamily arXiv:1407.0629 [nucl-th]}}.

\bibitem{pp22}
M.~Bleicher {\em et~al.}
  \href{http://dx.doi.org/10.1088/0954-3899/25/9/308}{{\em J.Phys.} {\bfseries
  G25} (1999) 1859--1896},
\href{http://arxiv.org/abs/hep-ph/9909407}{{\ttfamily arXiv:hep-ph/9909407
  [hep-ph]}}.

\bibitem{pp21}
S.~Bass {\em et~al.}
  \href{http://dx.doi.org/10.1016/S0146-6410(98)00058-1}{{\em
  Prog.Part.Nucl.Phys.} {\bfseries 41} (1998) 255--369},
\href{http://arxiv.org/abs/nucl-th/9803035}{{\ttfamily arXiv:nucl-th/9803035
  [nucl-th]}}.

\bibitem{pp24}
T.~Sjostrand, S.~Ask, J.~R. Christiansen, R.~Corke, N.~Desai, {\em et~al.}
  \href{http://dx.doi.org/10.1016/j.cpc.2015.01.024}{{\em Comput.Phys.Commun.}
  {\bfseries 191} (2015) 159--177},
\href{http://arxiv.org/abs/1410.3012}{{\ttfamily arXiv:1410.3012 [hep-ph]}}.

\bibitem{pp23}
W.~Ehehalt and W.~Cassing
\href{http://dx.doi.org/10.1016/0375-9474(96)00097-8}{{\em Nucl.Phys.}
  {\bfseries A602} (1996) 449--486}.

\bibitem{BeBe5}
I.~Tanihata {\em et~al.}
\href{http://dx.doi.org/10.1103/PhysRevLett.55.2676}{{\em Phys.\ Rev.\ Lett.}
  {\bfseries 55} (1985) 2676}.

\bibitem{BeBe6}
I.~Weimer
  \href{http://arxiv.org/abs/https://edms.cern.ch/file/1308546/1}{{\ttfamily
  https://edms.cern.ch/file/1308546/1}}. Bachelor Thesis, Ludwig-Maximilian
  University Munich.

\bibitem{BeBe7}
W.~Broniowski, M.~Rybczynski, and P.~Bozek
\href{http://dx.doi.org/10.1016/j.cpc.2008.07.016}{{\em Comput.Phys.Commun.}
  {\bfseries 180} (2009) 69}.

\bibitem{emil}
E.~Kaptur, {[NA61/SHINE} Collab.] {\em PoS} {\bfseries CPOD} (2014) ,
  \href{http://arxiv.org/abs/http://pos.sissa.it/archive/conferences/217/053/CPOD2014\_053.pdf}{{\ttfamily
  http://pos.sissa.it/archive/conferences/217/053/CPOD2014\_053.pdf}}.

\bibitem{BeBe9}
C.~Alt {\em et~al.}, {[NA49} Collab.]
\href{http://dx.doi.org/10.1140/epjc/s2005-02391-9}{{\em Eur.\ Phys.\ J.}
  {\bfseries C45} (2006) 343}.

\end{thebibliography}\endgroup
\bibliographystyle{na61Utphys}

\end{document}